\documentclass[11pt]{article}

\def\mytitle#1{\setcounter{equation}{0}
\setcounter{footnote}{0}
\begin{flushleft}\Large\textbf{#1}\end{flushleft}
\vspace{0.25cm}}
\def\myname#1{\leftline{{\large #1}}\vspace{-0.13cm}}
\def\myplace#1#2{\small\begin{flushleft}\textit{#1}\\
\texttt{#2}\end{flushleft}}

\usepackage{graphicx}
\usepackage{amsmath, amssymb, graphics}
\begin{document}

\mytitle{Study of Some Parameters of Modified Chaplygin Gas in
Galileon Gravity Theory from Observational Perspective}

\vskip0.2cm \myname{Chayan
Ranjit\footnote{chayanranjit@gmail.com}} \myplace{Department of
Mathematics, Seacom Engineering College, Howrah-711 302, India.}{}
\vskip0.2cm

\vskip0.2cm \myname{Prabir Rudra\footnote{prudra.math@gmail.com}}
\myplace{Department of Mathematics, Pailan College of Management
and Technology, Bengal Pailan Park, Kolkata-700 104, India.}{}

\vskip0.2cm \myname{Ujjal
Debnath\footnote{ujjaldebnath@gmail.com}} \myplace{Department of
Mathematics, Indian Institute of Engineering Science and
Technology, Shibpur, Howrah-711 103, India.}{} \vskip0.2cm

\begin{abstract}
We have assumed the FRW model of the universe in Galileon gravity,
which is filled with dark matter and Modified Chaplygin gas (MCG)
type dark energy. We present the Hubble parameter in terms of some
unknown parameters and observational parameters with the redshift
$z$. Some cosmological parameters are reconstructed and plots are
generated to study the nature of the model and its viability. It
is seen that the model is perfectly consistent with the present
cosmic acceleration. From \textit{observed Hubble data (OHD) set
or Stern data set} of 12 points, we have obtained the bounds of
the arbitrary parameters ($A,B$) \& ($A,C$) by minimizing the
$\chi^{2}$ test. Next due to joint analysis of \textit{Stern+BAO}
and \textit{Stern+BAO+CMB} observations, we have also obtained the
best fit values and the bounds of the parameters ($A,B$) \&
($A,C$) by fixing some other parameters. The best-fit values and
bounds of the parameters are obtained with 66\%, 90\% and 99\%
confidence levels for \textit{Stern, Stern+BAO and Stern+BAO+CMB}
joint analysis. Next we have also taken type Ia supernovae data
set (union 2 data set with 557 data points). The distance modulus
$\mu(z)$ against redshift $z$ for our theoretical MCG model in
Galileon gravity have been tested for the best fit values of the
parameters and the observed \textit{SNe Ia union 2 data} sample
and from this, we have concluded that our model is in agreement
with the union 2 sample data.
\end{abstract}

\section{Introduction}
Observational evidence strongly points to an accelerated expansion
of the Universe, but the physical origin of this acceleration is
still unknown. The observations include type Ia Supernovae and
Cosmic Microwave Background (CMB)
\cite{Perlmutter,Riess,Riess1,Bennet,Sperge} radiation. The
standard explanation invokes an unknown ``dark energy'' component
which has the property that positive energy density and negative
pressure. Observations indicate that dark energy occupies about
70\% of the total energy of the universe, and the contribution of
dark matter is $\sim$ 26\%. This accelerated expansion of the
universe has also been strongly confirmed by some other
independent experiments like Sloan Digital Sky Survey (SDSS)
\cite{Adel}, Baryonic Acoustic Oscillation (BAO)
\cite{Eisenstein}, WMAP data analysis \cite{Briddle,Spergel} etc.
Over the past decade, there have been many theoretical models for
mimicking the dark energy behaviors, such as the simplest (just)
cosmological constant in which the equation of state is
independent of the cosmic time and which can fit the observations
well. This model is the so-called $\Lambda$CDM, containing a
mixture of cosmological constant $\Lambda$ and cold dark matter
(CDM). However, two problems arise from this scenario, namely
``fine-tuning'' and the ``cosmic coincidence'' problems. In order
to solve these two problems, many dynamical dark energy models
were suggested, whose equation of state evolves with cosmic time.
The scalar field or quintessence \cite{Peebles,Cald} is one of the
most favored candidate of dark energy which produce sufficient
negative pressure to drive cosmic acceleration. In order to
alleviate the cosmological-constant problems and explain the
acceleration expansion, many dynamical dark energy models have
been proposed, such as K-essence, Tachyon, Phantom, quintom,
Chaplygin gas model, etc \cite{Arme,Sen,Cald1,Feng,Kamen}. Also
the interacting dark energy models including Modified Chaplygin
gas \cite{Debnath}, holographic dark energy model \cite{Cohen},
and braneworld model \cite{Sahni} have been proposed. Recently,
based on principle of quantum gravity, the agegraphic dark energy
(ADE) and the new agegraphic dark energy (NADE) models were
proposed by Cai \cite{Cai} and Wei et al \cite{Wei} respectively.
The theoretical models have been tally with the observations with
different data sets say TONRY, Gold sample data sets
\cite{Paddy1,Riess1,Tonry,Barris} etc. In Einstein's gravity, the
modified Chaplygin gas \cite{Debnath} best fits with the 3 year
WMAP and the SDSS data with the choice of parameters $A =0.085$
and $\alpha = 1.724$ \cite{Lu} which are improved constraints than
the previous ones $-0.35 < A < 0.025$ \cite{Jun}.\\

Another possibility is that general relativity is only accurate on
small scales and has to be modified on cosmological distances. One
of these is modified gravity theories. In this case, cosmic
acceleration would arise not from dark energy as a substance but
rather from the dynamics of modified gravity. Modified gravity
constitutes an interesting dynamical alternative to $\Lambda$CDM
cosmology in that it is also able to describe the current
acceleration in the expansion of our universe. One of the simplest
modified gravity is DGP brane-world model \cite{Dvali}. The other
alternative approach dealing with the acceleration problem of the
Universe is changing the gravity law through the modification of
action of gravity by means of using $f(R)$ gravity \cite{An,Noj0}
instead of the Einstein-Hilbert action. Some of these models, such
as $1/R$ and logarithmic models, provide an acceleration for the
Universe at the present time \cite{clif}. Other modified gravity
includes $f(T)$ gravity, $f(G)$ gravity, Gauss-Bonnet gravity,
Horava-Lifshitz gravity, Brans-Dicke gravity, etc
\cite{Yer,Noj,An1,Hora,Brans}.\\

Of late an infrared modification of classical gravitation was
proposed, as a generalization of the 4D effective theory in the
DGP model \cite{Nicolis1}. The theory considers a self-interaction
term of the form $\left(\nabla \phi\right)^{2}~^{\fbox{}}~\phi$ in
order to recover GR in high density regions. The most striking
feature of the theory is that it is invariant under the Galileon
shift symmetry, $\delta_{\mu}\phi\rightarrow
\delta_{\mu}\phi+c_{\mu}$, in the Minkowski background. Due to
this invariance the equation of motion remains a second order
differential equation, preventing the introduction of extra
degrees of freedom, which are usually associated with
instabilities. So we assume the FRW universe in Galileon gravity
model filled with the dark matter and the modified Chaplygin gas
(MCG) type dark energy. In \cite{Ranjit1} the observational
constraints of MCG was studied in RS II brane. In
\cite{Chakraborty1} the observational constraints of MCG was
studied in LQC. Moreover in \cite{Debnath2}  the parameter
constraints of MCG was studied in Einstein-Aether gravity.
Motivated by these we set to constrain the parameters of MCG in
Galileon gravity theory. The bounds on the parameters are to be
obtained using the observational data analysis mechanism. The
success of any dark energy or modified gravity model, depends
basically on its consistency with the observational data. This is
our basic motivation for the work. We reconstruct the hubble
parameter $H$ using the parameters of dark energy, dark matter and
modified gravity. Then we set up a comparison scenario between the
reconstructed $H$ ($H_{theoretical}$) and the values of $H$
obtained from observational data ($H_{observational}$). This is
accomplished by the procedure of chi-square test.

The basic concepts of Galileon gravity theory are presented in
section 2. The behaviour of some reconstructed cosmological
parameters is studied in section 3. The observational data
analysis tools in observed Hubble data (OHD) or $H(z)$-$z$
(Stern), OHD+BAO and OHD+BAO+CMB for $\chi^{2}$ minimum test will
be studied in section 4 and we will also investigate the bounds of
unknown parameters $(A,B)$ \& $(A,C)$ of MCG dark energy by fixing
other parameters. The best-fit values of the parameters are
obtained by 66\%, 90\% and 99\% confidence levels. The distance
modulus $\mu(z)$ against redshift $z$ for our theoretical model of
the MCG in Galileon gravity model for the best fit values of the
parameters and the observed SNe Ia union2 data sample. Finally the
paper ends with a discussion in section 5.

\section{\bf{Basic Equations and Solutions for MCG in Galileon
Gravity\\ Theory}}

The Galileon gravity theory is described by the action
\cite{Nicolis1,Deffayet1,Deffayet2,Chow1,Silva1}:
\begin{equation}\label{Lag}
S=\int d^{4} x \sqrt{-g}\left[\phi R- \frac{\omega}{\phi}
\left(\nabla \phi\right)^{2}+ f(\phi)^{\fbox{}}~\phi \left(\nabla
\phi\right)^{2}+{\cal L}_{m}\right]
\end{equation}
where $\phi$ is the Galileon field and the coupling function
$f(\phi)$ has dimension of length, $\left(\nabla
\phi\right)^{2}=g^{\mu\nu}\nabla_{\mu}\phi \nabla_{\nu}\phi$,~
$^{\fbox{}}~\phi=g^{\mu\nu}\nabla_{\mu}\nabla_{\nu}\phi$ and
${\cal L}_{m}$ is the matter Lagrangian. Variation of (1) with
respect to the metric $g_{\mu\nu}$ gives the Einstein's equations,

$$G_{\mu
\nu}=\frac{T_{\mu\nu}}{2\phi}+\frac{1}{\phi}\left(\nabla_{\mu}\nabla_{\nu}\phi-g_{\mu\nu}~^{\fbox{}}~\phi\right)
+\frac{\omega}{\phi^{2}}\left[\nabla_{\mu}\phi\nabla_{\nu}\phi-\frac{1}{2}g_{\mu\nu}\left(\nabla\phi\right)^{2}\right]$$
\begin{equation}
-\frac{1}{\phi}\left\{\frac{1}{2}g_{\mu\nu}\nabla_{\lambda}[f(\phi)\left(\nabla
\phi\right)^{2}]\nabla^{\lambda}\phi-\nabla_{\mu}[f(\phi)\left(\nabla
\phi\right)^{2}]\nabla_{\nu}\phi+f(\phi)\nabla_{\mu}\phi\nabla_{\nu}\phi~^{\fbox{}}~\phi\right\}
\end{equation}

For the Friedmann-Robertson-Walker background metric, the
Einstein's field eqns for Galileon gravity gives,
\begin{equation}
3H^{2}=\frac{\rho}{2\phi}-3HI+\frac{\omega}{2}I^{2}+\phi^{2}f(\phi)\left(3H-\frac{\alpha_{1}}{2}I\right)I^{3}
\end{equation}
and
\begin{equation}
-3H^{2}-2\dot{H}=\frac{p}{2\phi}+\dot{I}+I^{2}+2HI+\frac{\omega}{2}I^{2}-\phi^{2}f(\phi)\left(\dot{I}+\frac{2+\alpha_{1}}{2}I^{2}\right)I^{2}
\end{equation}
where $H(t)=\frac{\dot{a}}{a}$, $I(t)=\frac{\dot{\phi}}{\phi}$ and
$\alpha_{n}[\phi(t)]=\frac{d^{n}\ln f}{d \ln \phi^{n}}$ .
\vspace{5mm}

Here $\rho=\rho_{x}+\rho_{m}$ and $p=p_{x}+p_{m}$, where
$\rho_{m}$ and $p_{m}$ are the energy density and pressure of the
dark matter with the equation of state given by $p_{m} =
w_{m}\rho_{m}$ and $\rho_{x}$, $p_{x}$ are respectively the energy
density and pressure contribution of some dark energy. Here we
consider the universe filled with Modified Chaplygin Gas (MCG).
The equation of state (EOS) of MCG is given by \cite{Debnath}

\begin{equation}
p_{x} = A\rho_{x}- \frac{B}{\rho_{x}^{\alpha}},~~ B >0 ,~~0\leq
\alpha \leq 1
\end{equation}
We also consider the dark matter and and the dark energy are
separately conserved and the conservation equations of dark matter
and dark energy (MCG) are given by
\begin{equation}
\dot{\rho}_{m}+3H(\rho_{m}+p_{m})=0
\end{equation}
and
\begin{equation}
\dot{\rho}_{x}+3H(\rho_{x}+p_{x})=0
\end{equation}

From first conservation equation (6) we have the solution of
$\rho_{m}$ as
\begin{equation}
\rho_{m}=\rho_{m0}(1+z)^{3(1+w_m)}
\end{equation}

From the conservation equation (7) we have the solution of the
energy density as
\begin{equation}
\rho_{x}=\left[\frac{B}{A+1}+C(1+z)^{3(\alpha+1)(A+1)}\right]^{\frac{1}{\alpha+1}}
\end{equation}
where $C$ is the integrating constant, $z=\frac{1}{a}-1$ is the
cosmological redshift (choosing $a_{0}=1$) and the first constant
term can be interpreted as the contribution of dark energy. So the
above equation can be written as
\begin{equation}
\rho_{x}=\rho_{x0}\left[\frac{B}{(1+A)C+B}+\frac{(1+A)C}{(1+A)C+B}(1+z)^{3(\alpha+1)(A+1)}\right]^{\frac{1}{\alpha+1}}
\end{equation}
where $\rho_{x0}$ is the present value of the dark energy
density.\\

\section{\bf{Behaviour of Some Reconstructed Cosmological Parameters}}

\subsection{\bf{Deceleration parameter}}

From the solution of MCG (eqn. 10) and defining the dimensionless
density parameters $\Omega_{m0}=\frac{\rho_{m0}}{3 H_{0}^{2}}$ and
$\Omega_{x0}=\frac{\rho_{x0}}{3 H_{0}^{2}}$ and for simplicity
choosing $f(\phi)=f_{0}\phi^{n}$ and $\phi=\phi_{0}a^{m}$
($f_{0}>0,~\phi_{0}>0,~m>0,~n>0$), we have the expression for
Hubble parameter $H$ in terms of redshift parameter $z$ as follows
(from eq. 3):

$$H(z)=\frac{1}{2 \left(6 f_{0} m^3 \phi_{0}^{3+n}-f_{0} m^4 n \phi_{0}^{3+n}\right)}\left[-(m^2
\omega-6m-6) (1+z)^{m (2+n)} \phi_{0}\right.$$

$$\left.+ \left\{\left((m^{2}\omega-6m-6)(1+z)^{m (2+n)} \phi_
{0}\right)^2-f_{0} m^4 n \phi_ {0}^{3+n}-12H_{0}^{2}
\left((1+z)^{3+m (3+n)+3 w_m} \Omega_
{m_{0}}+\right.\right.\right.$$

\begin{equation}
\left.\left.\left. (1+z)^{m (3+n)}\left(\frac{B}{B+(1+A)
C}+\frac{(1+A) C (1+z)^{3 (1+A) (1+\alpha )}}{B+(1+A)
C}\right)^{\frac{1}{1+\alpha }} \Omega_ {x_{0}}\right) 6 f_{0} m^3
\phi_ {0}^{3+n}\right\}^{\frac{1}{2}}\right]
\end{equation}

\begin{figure}
~~~~~~~~~~~~~~~~~~~~~~~~~~~~~\includegraphics[height=2.5in]{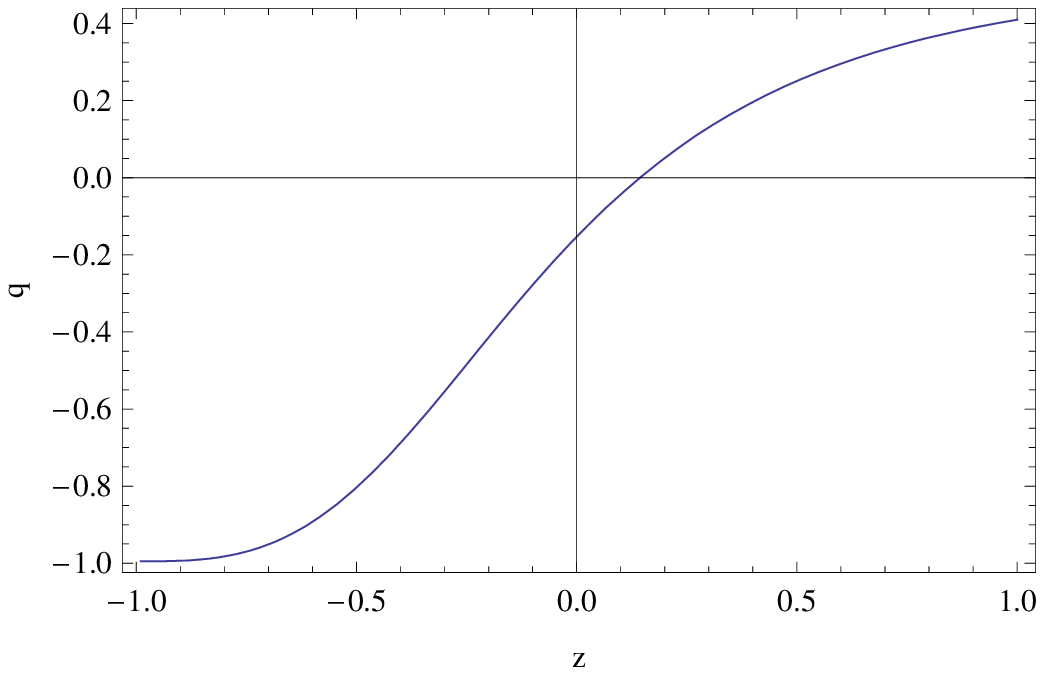}\\
\vspace{1cm}
~~~~~~~~~~~~~~~~~~~~~~~~~~~~~~~~~~~~~~~~~~~~~~~~~~~~~~~~~~~Fig.1~~~~~~~~~~~~~~~~~~~~~~~~~~~~~~~~~~~~~~~~~~~~~~~~~~~~~~~~~\\
\vspace{1mm}

Fig.1 shows the variation of deceleration parameter against the
redshift. \vspace{1cm}

\end{figure}

The study of deceleration parameter is very important for any
cosmological model, considering the recent cosmic acceleration.
The expression for the parameter is given by,
\begin{equation}
q=-1-\frac{\dot{H}}{H^{2}}=-1+\frac{(1+z)}{H} \frac{dH}{dz}
\end{equation}

Here we reconstruct the deceleration parameter, $q$ for the model
under consideration. Using equation (11) and (12), the expression
for $q$ is obtained as,

$$q=-1+\left((1+z)\left(-12m(2+n)(1+z)^{-1+m(2+n)}\phi_{0}-12
m^2(2+n)(1+z)^{-1+m(2+n)}\phi_{0}\right.\right.$$

$$\left.\left.+2m^3(2+n)\omega(1+z)^{-1+m (2+n)}\phi_{0}+\left(2m(2+n)\left(6+6
m-m^2\omega\right)^2 (1+z)^{-1+2
m(2+n)}\phi_{0}^2\right.\right.\right.$$

$$\left.\left.\left.-4f_{0}m^3(-6+mn)(1+z)^{m(3+n)}\left(-3(1+z)^{3w_{m}}
\rho_{m0}-6z(1+z)^{3w_{m}}\rho_{m0}-3z^2(1+z)^{3w_{m}}\rho_{m0}\right.\right.\right.\right.$$

$$\left.\left.\left.\left.-(m(3+n)+3w_{m})(1+z)^{-1+3w_{m}}\rho_{m0}-3(m
(3+n)+3w_{m})z(1+z)^{-1+3w_{m}}\rho_{m0}\right.\right.\right.\right.$$

$$\left.\left.\left.\left.-3(m(3+n)+3w_{m})z^2(1+z)^{-1+3w_{m}}\rho_{m0}-(m
(3+n)+3w_{m})z^3(1+z)^{-1+3w_{m}}\rho{m0}\right.\right.\right.\right.$$

$$\left.\left.\left.\left.-\frac{m(3+n)\left(\frac{B+(1+A)C(1+z)^{3(1+A)
(1+\alpha)}}{B+C+AC}\right)^{\frac{1}{1+\alpha}}
\rho_{x0}}{1+z}-\frac{3(1+A)^2C
\left(\frac{B+(1+A)C(1+z)^{3(1+A)(1+\alpha)}}{B+C+A
C}\right)^{\frac{1}{1+\alpha}}\rho_{x0}}{(1+z)\left(C+AC+B
(1+z)^{-3(1+A)(1+\alpha)}\right)}\right)
\phi_{0}^{3+n}\right)\right.\right.$$

$$\left.\left.\left(\surd\left(\left(6+6 m-m^2
\omega\right)^2(1+z)^{2m(2+n)}\phi_{0}^2+4f_{0}m^3(-6+m n)
(1+z)^{m(3+n)}\left((1+z)^{3+3
w_{m}}\rho_{m0}\right.\right.\right.\right.\right.$$

$$\left.\left.\left.\left.\left.+\left(\frac{B+(1+A)C(1+z)^{3(1+A)
(1+\alpha)}}{B+C+AC}\right)^{\frac{1}{1+\alpha}}
\rho_{x0}\right)\phi_{0}^{3+n}\right)\right)\right)\right)4
\left(-6(1+z)^{m(2+n)}\phi_{0}-6m(1+z)^{m(2+n)}\phi_{0}\right.$$

$$\left.+m^2 \omega(1+z)^{m(2+n)}\phi_{0}+\surd \left(\left(6+6
m-m^2\omega\right)^2(1+z)^{2m(2+n)}\phi_{0}^2+4f_{0}m^3(-6+m n)
(1+z)^{m(3+n)}\right.\right.$$

\begin{equation}
\left.\left.\left((1+z)^{3+3 w_{m}} \rho_{m0}+\left(\frac{B+(1+A)
C (1+z)^{3 (1+A) (1+\alpha )}}{B+C+A C}\right)^{\frac{1}{1+\alpha
}} \rho_{x0}\right) \phi_{0}^{3+n}\right)\right)
\end{equation}

\subsection{\bf{EoS parameter}}
The Equation of state (EoS) parameter, $w$ determines the nature
of matter content of the universe. Its value gives an idea about
the era of the universe. $w=1$ and $w=1/3$ predicts the stiff
fluid and the radiation era respectively. $w=0$ gives the dust
era. $w=-1/3$ corresponds to quintessence (dark energy). $w=-1$
and $w<-1$ represent the $\Lambda CDM$ and phantom era
respectively. The EoS parameter can be obtained as
\begin{equation}
w=\frac{2q-1}{3}
\end{equation}
Here we reconstruct it for the given model, and plot it against
the redshift parameter in fig.2.
\begin{figure}
~~~~~~~~~~~~~~~~~~~~\includegraphics[height=2.5in]{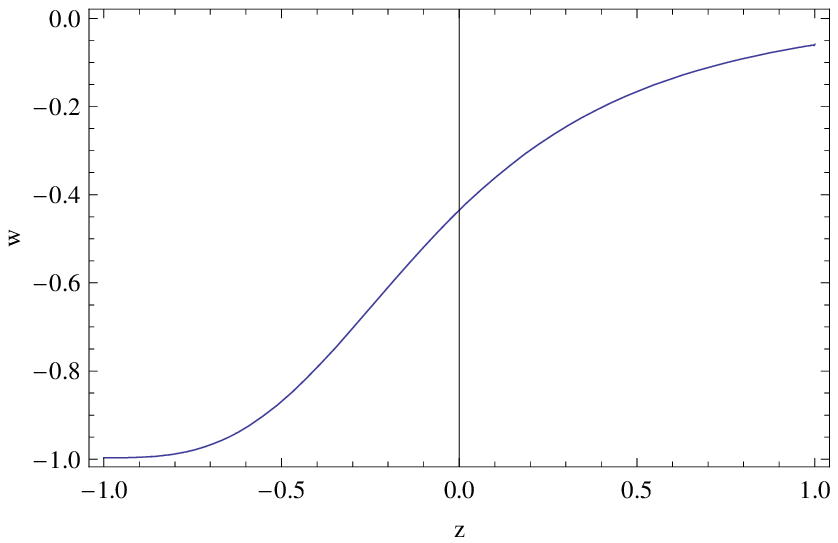}\\
\vspace{1cm}
~~~~~~~~~~~~~~~~~~~~~~~~~~~~~~~~~~~~~~~~~~~~~~~~Fig.2~~~~~~~~~~~~~~~~~~~~~~~~~~~~~~~~~~~~~~~~~~~~~~~~~~~~~~~~~\\
\vspace{1mm}

Fig.2 gives the plot of the EoS parameter against the redshift
parameter. \vspace{1cm}

\end{figure}

\subsection{\bf{Statefinder parameters}}
In order to distinguish between the numerous dark energy models,
Sahni et al in 2003 \cite{Sahni1} proposed a cosmological
diagnostic pair $\{r,s\}$ which is known as as statefinder
parameters. Since the two parameters are derived from the cosmic
scale factor alone, they are dimensionless and geometrical in
nature. The diagnostic pair is defined as follows:

\begin{equation}
r=1+3\frac{\dot{H}}{H^{2}}+\frac{\ddot{H}}{H^{3}} ~~\text{and} ~~
s=\frac{r-1}{3(q-\frac{1}{2})}
\end{equation}
Clear difference in trajectories are found when different dark
energy models are examined in the $r-s$ plane, thus making the
pair extremely important in the study of dark energy. Here we
reconstruct the diagnostic pair for the given model using equation
(11) and (15), and the simulate their nature in figs. 3, 4 and
5.\\

\begin{figure}
\includegraphics[height=2in]{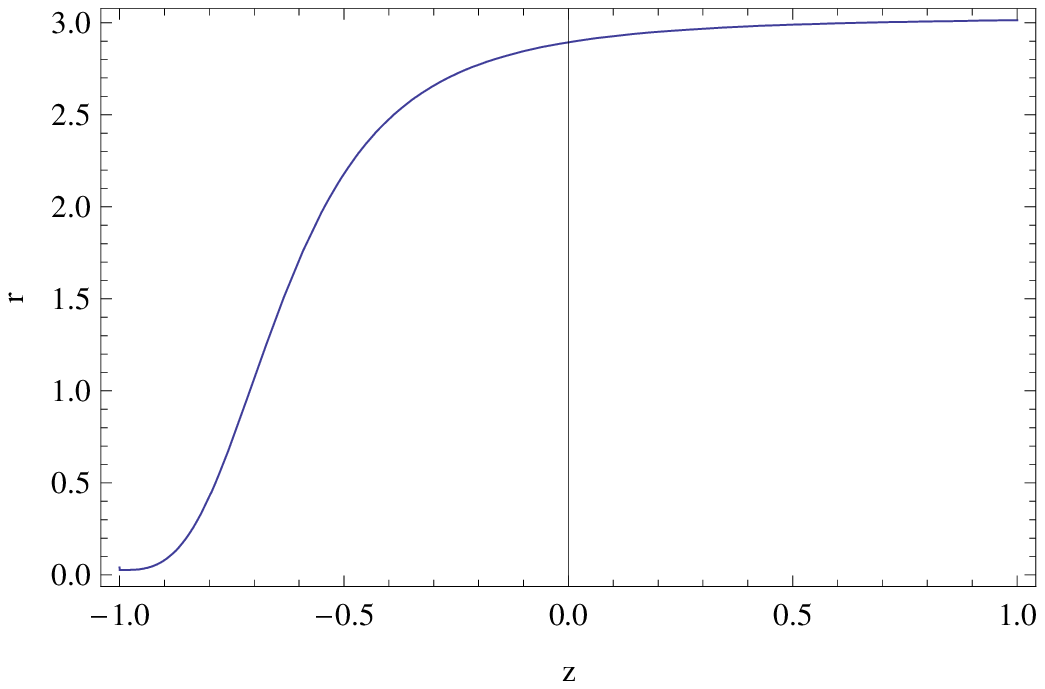}~~\includegraphics[height=2in]{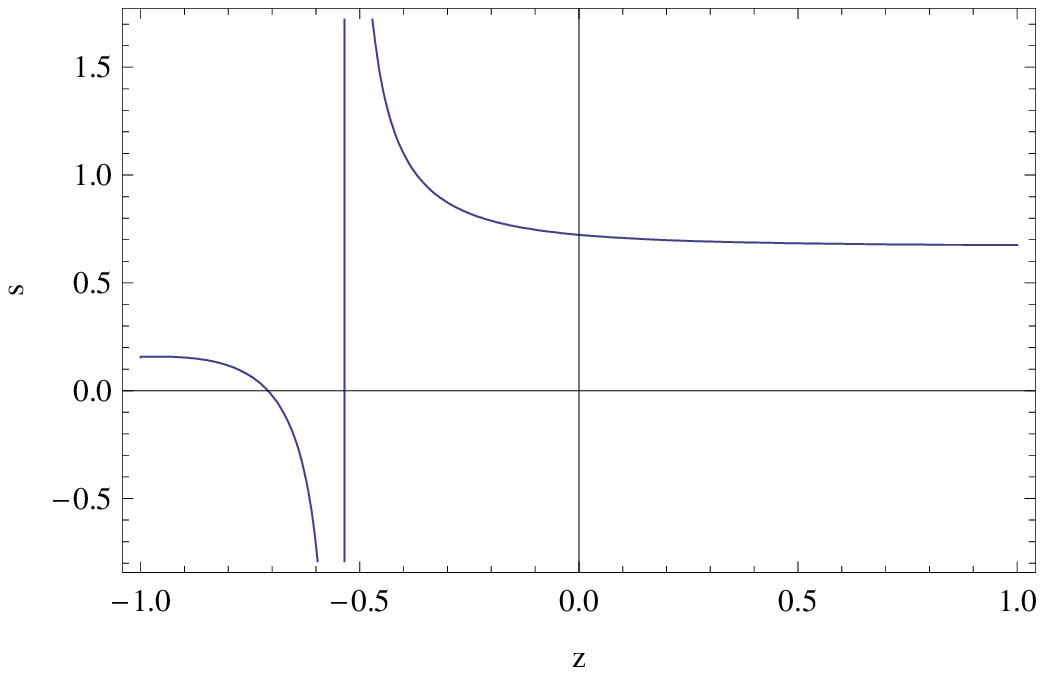}\\
\vspace{2mm}
~~~~~~~~~~~~~~~~~~~~~~~~~~~~~Fig.3~~~~~~~~~~~~~~~~~~~~~~~~~~~~~~~~~~~~~~~~~~~~~~~~~~~~~~~~~~~Fig.4\\
\vspace{2mm} ~~~~~~~~~~~~~~~~~~~~~~~~~~~~~~\includegraphics[height=2.5in]{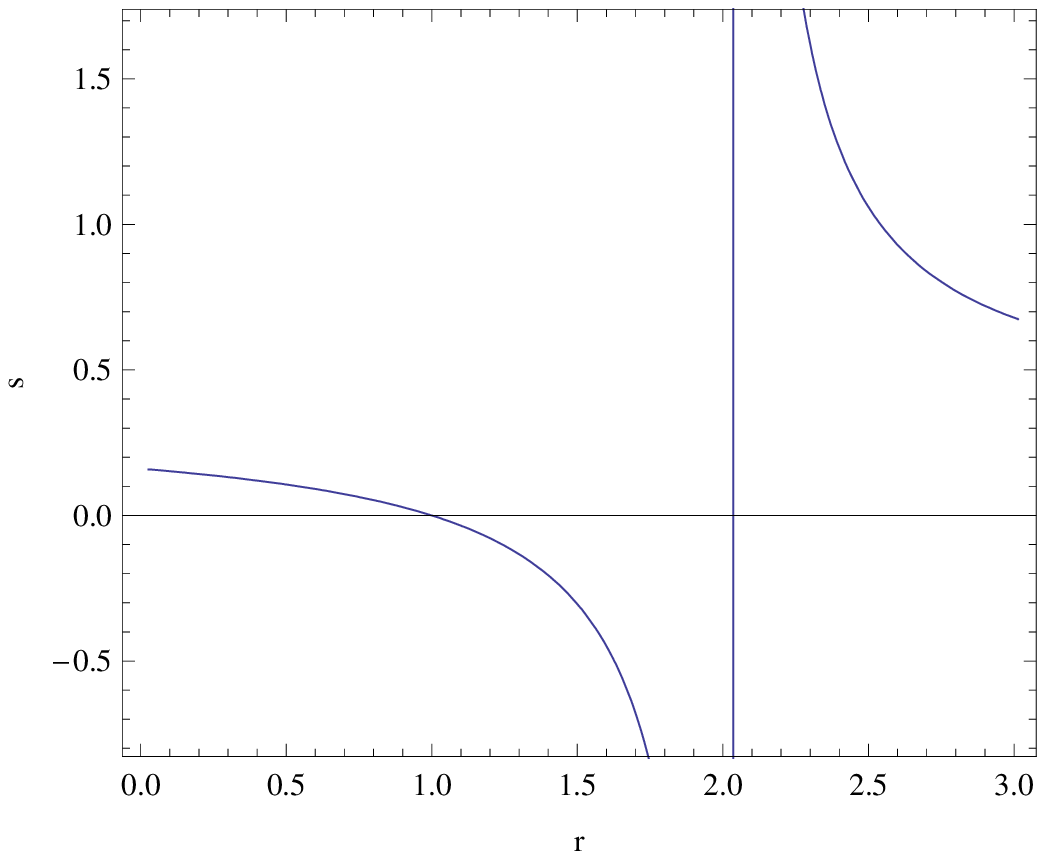}\\
\vspace{2mm}
~~~~~~~~~~~~~~~~~~~~~~~~~~~~~~~~~~~~~~~~~~~~~~~~~~~~~~~~~~~~~Fig.5~~~~~~~~~~~~~~~~~~~~~~~~~~~\\
\vspace{1mm}

Fig.3-4 show the plots of statefinder parameters against redshift.\\
Fig.5 gives the trajectories in $r$-$s$ plane.\\
\vspace{1mm}

\end{figure}

In the next section, we shall investigate some bounds of the
parameters in Galileon gravity by observational data fitting. The
parameters are determined by $H(z)$-$z$ (Stern) or OHD, OHD+BAO
and OHD+BAO+CMB joint data analysis \cite{Wu1,Paul,Paul1}. We
shall use the $\chi^{2}$ minimization technique (statistical data
analysis) to get the constraints of the parameters of MCG in Galileon gravity model.\\

\section{\bf{Observational Data Analysis}}

From eqn. (11), we see that $H(z)$ contains the unknown parameters
like $A,~B,~C$, $\Omega_{m0}$, $\Omega_{x0}$, $\alpha$, $n$, $m$,
$\omega$, $w_m$, $f_0$, $\phi_0$. Now the relation between two
parameters will be obtained by fixing the other parameters and by
using observational data set. Eventually the bounds of the
parameters will be obtained by using this observational data
analysis mechanism.

\[
\begin{tabular}{|c|c|c|}
\hline
  ~~~~~~$z$ ~~~~& ~~~~$H(z)$ ~~~~~& ~~~~$\sigma(z)$~~~~\\
  \hline
  0 & 73 & $\pm$ 8 \\
  0.1 & 69 & $\pm$ 12 \\
  0.17 & 83 & $\pm$ 8 \\
  0.27 & 77 & $\pm$ 14 \\
  0.4 & 95 & $\pm$ 17.4\\
  0.48& 90 & $\pm$ 60 \\
  0.88 & 97 & $\pm$ 40.4 \\
  0.9 & 117 & $\pm$ 23 \\
  1.3 & 168 & $\pm$ 17.4\\
  1.43 & 177 & $\pm$ 18.2 \\
  1.53 & 140 & $\pm$ 14\\
  1.75 & 202 & $\pm$ 40.4 \\ \hline
\end{tabular}
\]
{\bf Table 1:} The Hubble parameter $H(z)$ and the standard error
$\sigma(z)$ for different values of redshift $z$.

\subsubsection{Analysis with Stern ($H(z)$-$z$) Data Set}

\begin{figure}
\includegraphics[height=3in]{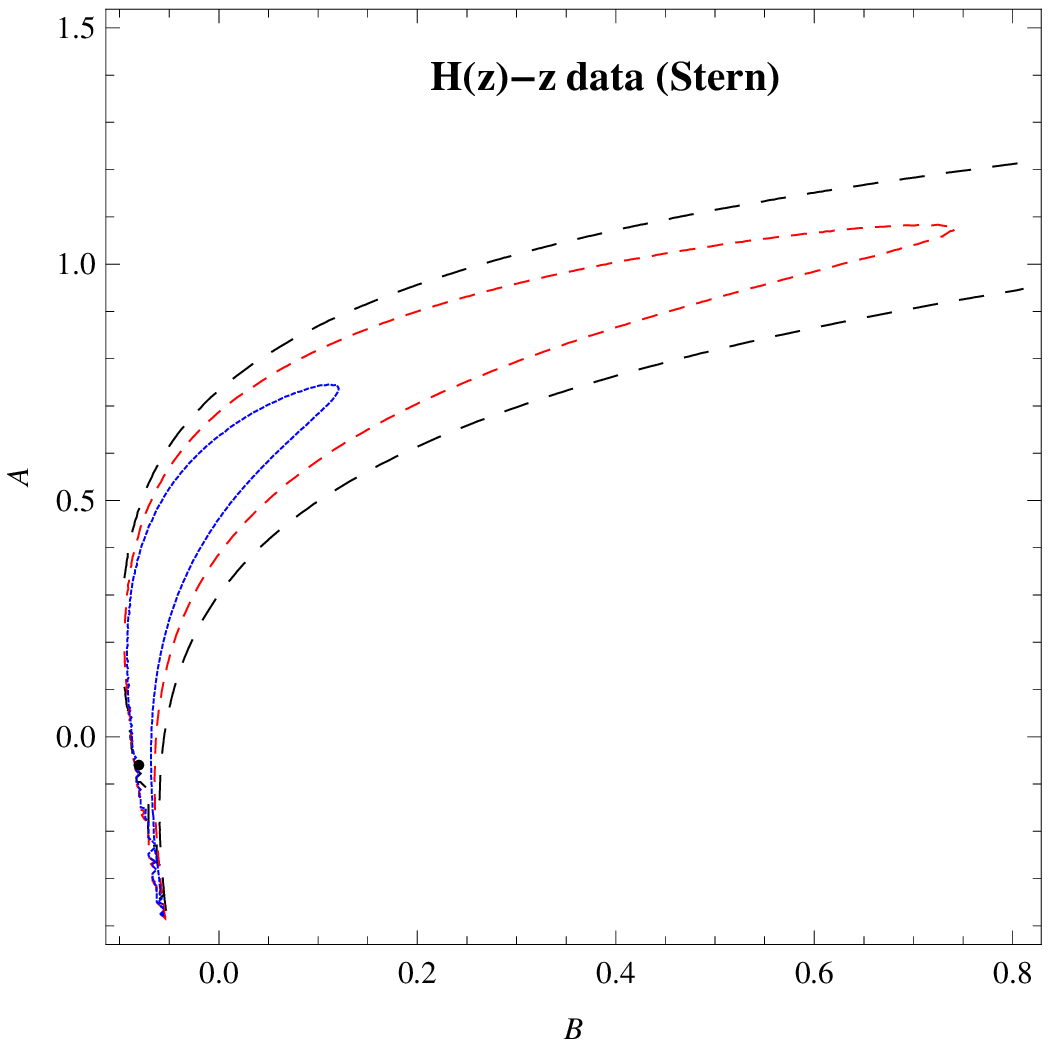}~~~~~~\includegraphics[height=3in]{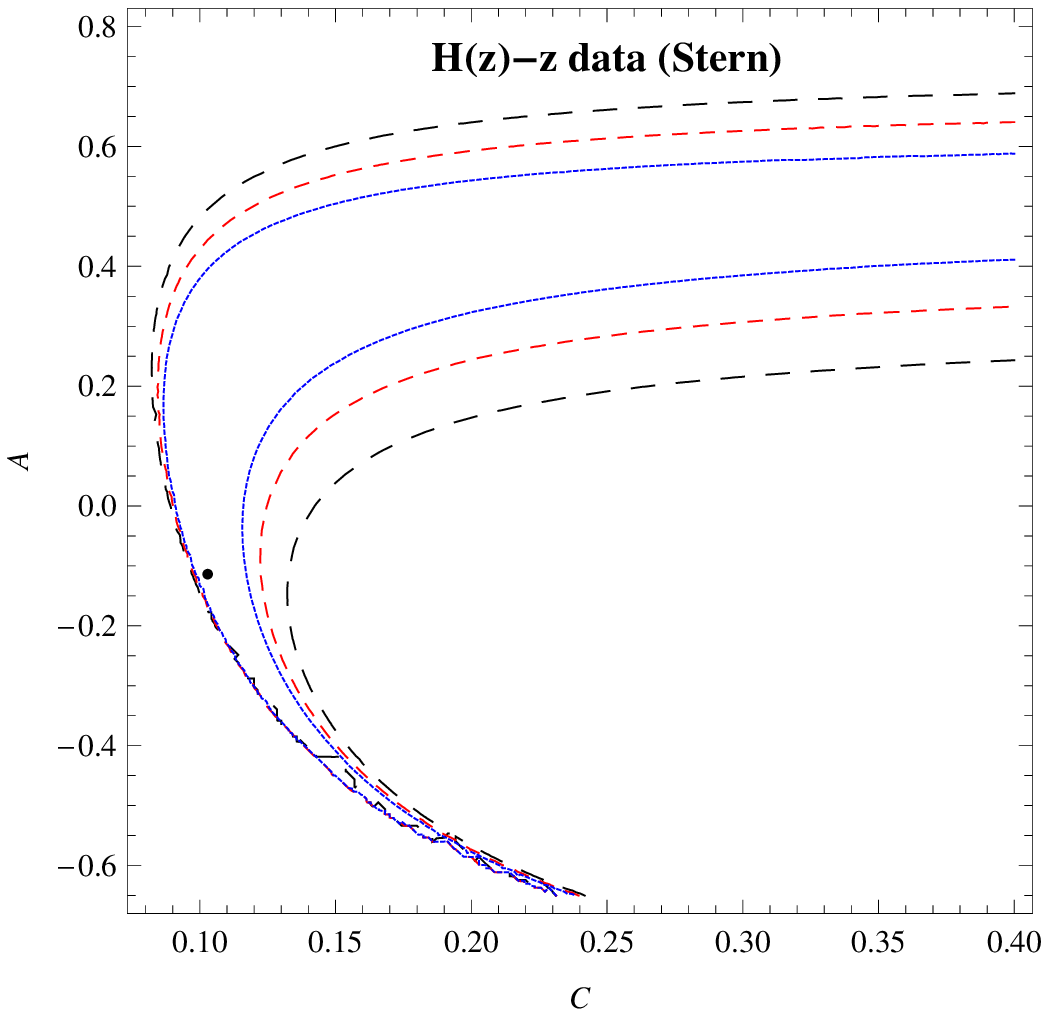}\\
\vspace{2mm}
~~~~~~~~~~~~~~~~~~~Fig.6~~~~~~~~~~~~~~~~~~~~~~~~~~~~~~~~~~~~~~~~~~~~~~~~~~~~~~~~~~~~~~~Fig.7\\
\vspace{1mm}

Fig.6 \& 7 show the variations of $A$ with $B$ and $A$ with $C$
respectively for different confidence levels in Galileon gravity
with  modified Chaplygin gas. The 66\% (solid, blue), 90\%
(dashed, red) and 99\% (dashed, black) contours are plotted in
these figures for the $H(z)$-$z$ (Stern) analysis. \vspace{1cm}

\end{figure}

We analyze the model, using observed value of Hubble parameter at
different redshifts (twelve data points) listed in observed Hubble
data by Stern et al \cite{Stern1}. The Hubble parameter $H(z)$ and
the standard error $\sigma(z)$ for different values of redshift
$z$ are given in Table 1. Since we are using testing of
hypothesis, so before proceeding, we form our null and alternate
hypothesis which are given below.\\

\textbf{Null Hypothesis: ~~~~~~~~~~~~~~~~~~~~$H_{0}$: $H_{theoretical}=H_{observational}$}\\

\textbf{Alternative Hypothesis:~~~~~~~~~~~~$H_{1}$:
$H_{theoretical}\neq
H_{observational}$}\\

Here we test the null hypothesis $H_{0}$ against the alternative
hypothesis $H_{1}$. For this purpose we first form the $\chi^{2}$
statistics as a sum of standard normal distribution as follows:

\begin{equation}
{\chi}_{Stern}^{2}=\sum\frac{(H(z)-H_{obs}(z))^{2}}{\sigma^{2}(z)}
\end{equation}

where $H(z)$ and $H_{obs}(z)$ are theoretical and observational
values of Hubble parameter at different redshifts respectively.
Here, $H_{obs}$ is a nuisance parameter and can be safely
marginalized. We consider the present value of Hubble parameter
$H_{0}$ = 72 $\pm$ 8 Kms$^{-1}$ Mpc$^{-1}$ and a fixed prior
distribution. Here we shall determine best fit value of the
parameters ($A,B$)\&($A,C$) by minimizing the above distribution
${\chi}_{Stern}^{2}$ and fixing the other unknown parameters with
the help of Stern data. We now plot the graph for different
confidence levels. In early stage the Chaplygin Gas follow the
equation of state $p=A\rho$ where $A\le 1$. So, as per our
theoretical model the two parameters should satisfy the two
inequalities $A\le 1$ and $B>0$. Now our best fit analysis with
Stern observational data support the theoretical range of the
parameters. The 66\% (solid, blue), 90\% (dashed, red) and 99\%
(dashed, black) contours are plotted in figures 6 and 7 for
$\Omega_{m0}=0.0014$, $\Omega_{x0}=0.0014$, $\alpha=0.001$,
$n=0.5$,$m=10$, $\omega=-3$, $w_m=0.03$, $f_0=0.01$,
$\phi_0=0.01$. The best fit values of ($A,B$) \& ($A,C$) are
tabulated in Table 2.

\[
\begin{tabular}{|c|c|c|c|}
\hline
  ~~~~~~$A$ ~~~~~& ~~~~~~~$B$ ~~~~~&~~~~~$\chi^{2}_{min}$~~~~~~\\
  \hline
  -0.0597946& -0.0805254 &  333.628 \\
    \hline
     ~~~~~~$A$ ~~~~~~~~& ~~~$C$~~~~~&~~~~~$\chi^{2}_{min}$~~~~~~\\
  \hline
  -0.115326&  0.102968 & 332.727  \\
  \hline
\end{tabular}
\]
{\bf Table 2:} $H(z)$-$z$ (Stern): The best fit values of $A$ with
$B$ and $C$ for the minimum values of $\chi^{2}$.

\subsubsection{Joint Analysis with Stern $+$ BAO Data Sets}

The method of joint analysis, the Baryon Acoustic Oscillation
(BAO) peak parameter value has been proposed by \cite{Eisenstein}
and we shall use their approach. Sloan Digital Sky Survey (SDSS)
survey  is one of the first redshift survey by which the BAO
signal has been directly detected at a scale $\sim$ 100 MPc. The
said analysis is actually the combination of angular diameter
distance and Hubble parameter at that redshift. This analysis is
independent of the measurement of $H_{0}$ and not containing any
particular dark energy. Here we examine the parameters $B$ and $C$
for Chaplygin gas model from the measurements of the BAO peak for
low redshift (with range $0<z<0.35$) using standard $\chi^{2}$
analysis. The error is corresponding to the standard deviation,
where we consider Gaussian distribution. Low-redshift distance
measurements is a lightly dependent on different cosmological
parameters, the equation of state of dark energy and have the
ability to measure the Hubble constant $H_{0}$ directly. The BAO
peak parameter may be defined by

\begin{equation}
{\cal
A}=\frac{\sqrt{\Omega_{m}}}{E(z_{1})^{1/3}}\left(\frac{1}{z_{1}}~\int_{0}^{z_{1}}
\frac{dz}{E(z)}\right)^{2/3}
\end{equation}
Here $E(z)=H(z)/H_{0}$ is the normalized Hubble parameter, the
redshift $z_{1}=0.35$ is the typical redshift of the SDSS sample
and the integration term is the dimensionless comoving distance to
the to the redshift $z_{1}$ The value of the parameter ${\cal A}$
for the flat model of the universe is given by ${\cal A}=0.469\pm
0.017$ using SDSS data \cite{Eisenstein} from luminous red
galaxies survey. Now the $\chi^{2}$ function for the BAO
measurement can be written as

\begin{equation}
\chi^{2}_{BAO}=\frac{({\cal A}-0.469)^{2}}{(0.017)^{2}}
\end{equation}

Now the total joint data analysis (Stern+BAO) for the $\chi^{2}$
function may be defined by

\begin{equation}
\chi^{2}_{total}=\chi^{2}_{Stern}+\chi^{2}_{BAO}
\end{equation}

According to our analysis the joint scheme(Stern+BAO) gives the
best fit values of ($A,B$) \& ($A,C$) in Table 3. Finally we draw
the contours for the 66\% (solid,blue), 90\% (dashed, red) and
99\%(dashed, black) confidence limits depicted in figures $8$ and
$9$ for $\Omega_{m0}=0.0014$, $\Omega_{x0}=0.0014$,
$\alpha=0.001$, $n=0.5$,$m=10$, $\omega=-3$, $w_m=0.03$,
$f_0=0.01$,
$\phi_0=0.01$.\\

\[
\begin{tabular}{|c|c|c|c|}
\hline
  ~~~~~~$A$ ~~~~~& ~~~~~~~$B$ ~~~~~&~~~~~$\chi^{2}_{min}$~~~~~~\\
  \hline
  1.95212 & 7.6453 &  373.061 \\
    \hline
     ~~~~~~$A$ ~~~~~~~~& ~~~$C$~~~~~&~~~~~$\chi^{2}_{min}$~~~~~~\\
  \hline
  2.0324 &  0.0763776 & 372.895  \\
  \hline
\end{tabular}
\]
{\bf Table 3:} $H(z)$-$z$ (Stern)+ BAO: The best fit values of $A$
with $B$ and $C$ for the minimum values of $\chi^{2}$.

\begin{figure}
\includegraphics[height=3in]{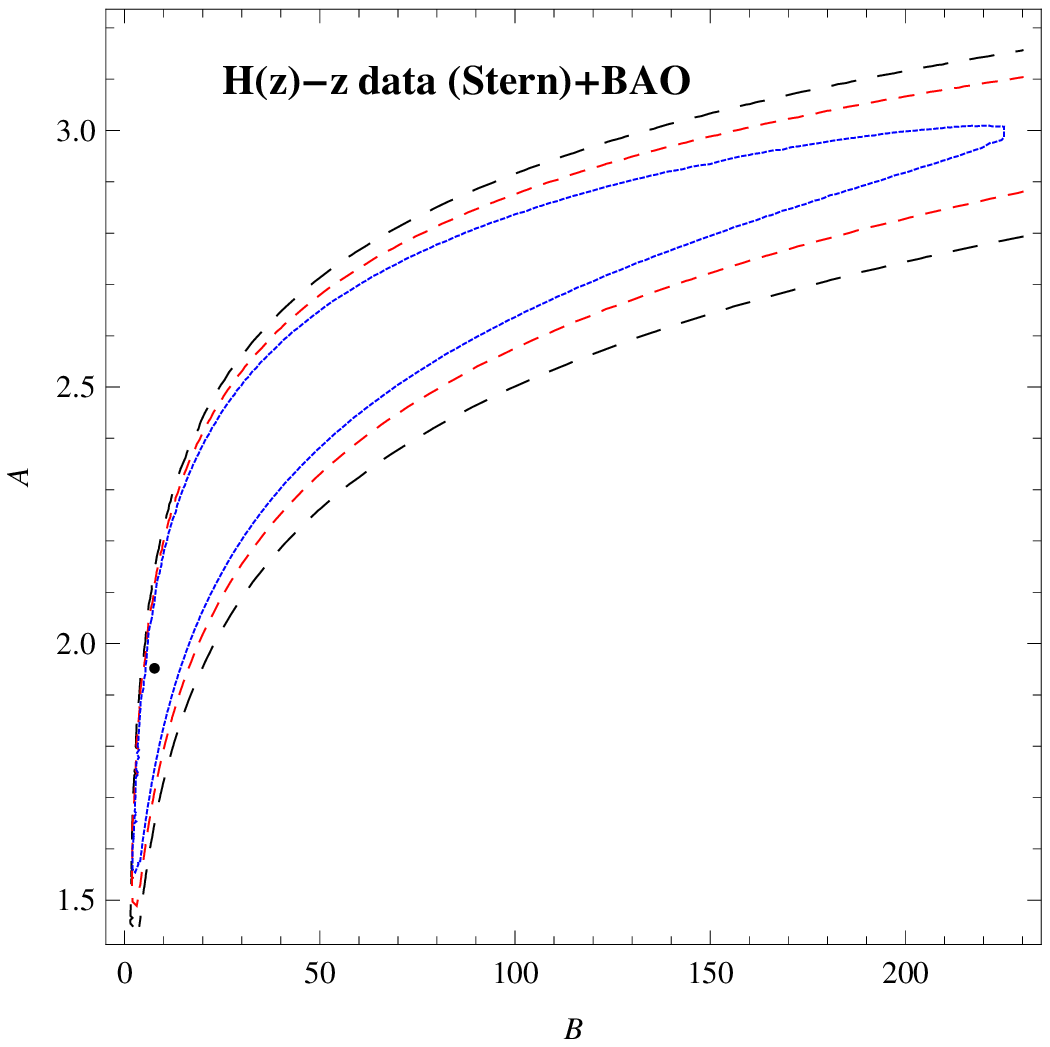}~~~~\includegraphics[height=3in]{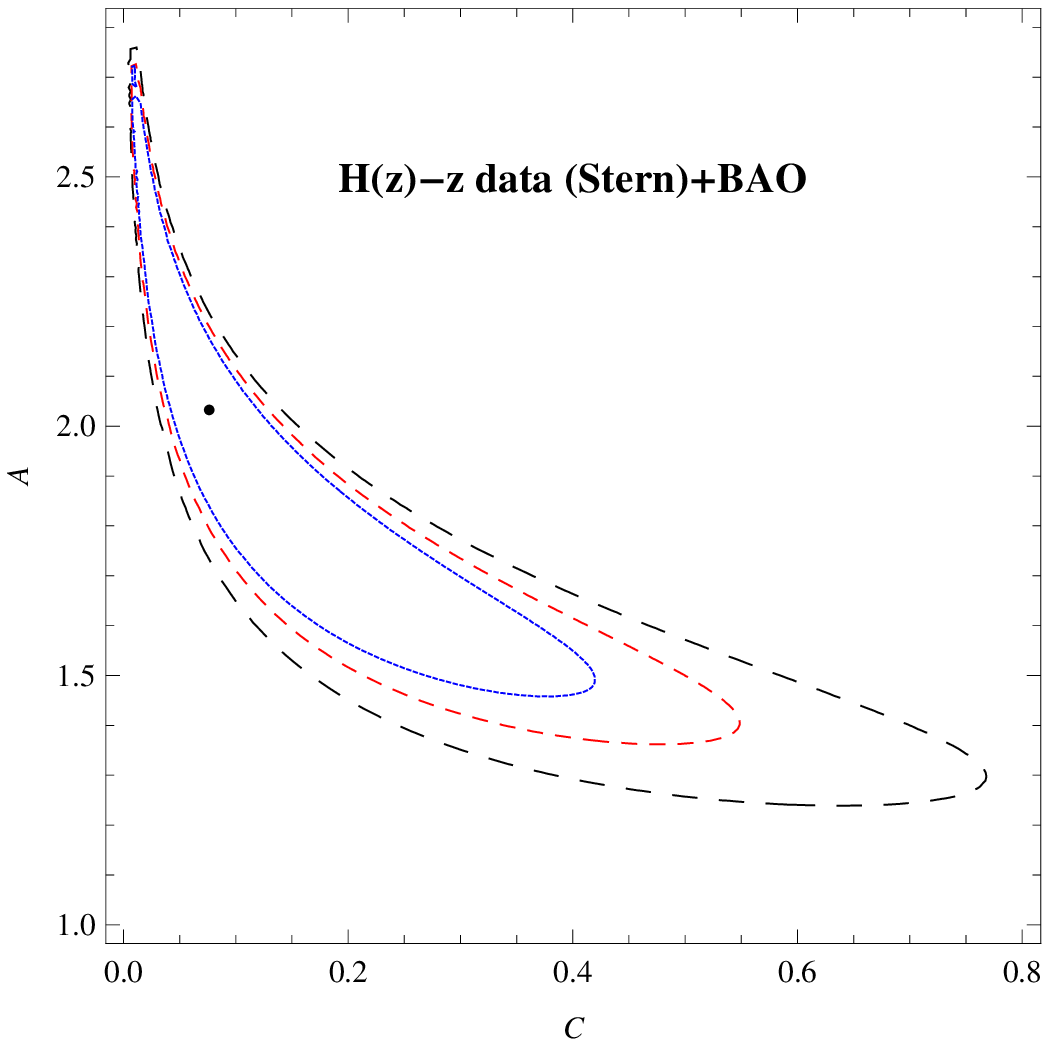}\\
\vspace{2mm}
~~~~~~~~~~~~~~~~~~~~~~~~~~~~Fig.8~~~~~~~~~~~~~~~~~~~~~~~~~~~~~~~~~~~~~~~~~~~~~Fig.9\\
\vspace{1mm}

Fig.8 \& 9 show the variations of $A$ with $B$ and $A$ with $C$
respectively for different confidence levels in Galileon gravity
with  modified Chaplygin gas. The 66\% (solid, blue), 90\%
(dashed, red) and 99\% (dashed, black) contours are plotted in
these figures for the $H(z)$-$z$ (Stern)+BAO joint analysis.
\vspace{1cm}

\end{figure}

\subsubsection{Joint Analysis with Stern $+$ BAO $+$ CMB Data Sets}

One interesting geometrical probe of dark energy can be determined
by the angular scale of the first acoustic peak through angular
scale of the sound horizon at the surface of last scattering which
is encoded in the CMB power spectrum Cosmic Microwave Background
(CMB) shift parameter is defined by \cite{Bond1, Efstathiou1,
Nessaeris1}. It is not sensitive with respect to perturbations but
are suitable to constrain model parameter. The CMB power spectrum
first peak is the shift parameter which is given by

\begin{equation}
{\cal R}=\sqrt{\Omega_{m}} \int_{0}^{z_{2}} \frac{dz}{E(z)}
\end{equation}

where $z_{2}$ is the value of redshift at the last scattering
surface. From WMAP7 data of the work of Komatsu et al
\cite{Komatsu1} the value of the parameter has obtained as ${\cal
R}=1.726\pm 0.018$ at the redshift $z=1091.3$. Now the $\chi^{2}$
function for the CMB measurement can be written as

\begin{equation}
\chi^{2}_{CMB}=\frac{({\cal R}-1.726)^{2}}{(0.018)^{2}}
\end{equation}

Now when we consider three cosmological tests together, the total
joint data analysis (Stern+BAO+CMB) for the $\chi^{2}$ function
may be defined by

\begin{equation}
\chi^{2}_{TOTAL}=\chi^{2}_{Stern}+\chi^{2}_{BAO}+\chi^{2}_{CMB}
\end{equation}
Now the best fit values of ($A,B$) \& ($A,C$) for joint analysis
of BAO and CMB with Stern observational data support the
theoretical range of the parameters given in Table 4. The 66\%
(solid, blue), 90\% (dashed, red) and 99\% (dashed, black)
contours are plotted in figures 10 and 11 for
$\Omega_{m0}=0.0014$, $\Omega_{x0}=0.0014$, $\alpha=0.001$,
$n=0.5$,$m=10$, $\omega=-3$, $w_m=0.03$, $f_0=0.01$,
$\phi_0=0.01$.

\[
\begin{tabular}{|c|c|c|c|}
\hline
  ~~~~~~$A$ ~~~~~& ~~~~~~~$B$ ~~~~~&~~~~~$\chi^{2}_{min}$~~~~~~\\
  \hline
  -0.0099825 & -27.5635 &  10121.654 \\
    \hline
     ~~~~~~$A$ ~~~~~~~~& ~~~$C$~~~~~&~~~~~$\chi^{2}_{min}$~~~~~~\\
  \hline
  -0.0100278 &  0.020933 & 10121.739  \\
  \hline
\end{tabular}
\]
{\bf Table 4:} $H(z)$-$z$ (Stern) + BAO + CMB : The best fit
values of $A$ with $B$ and $C$ for the minimum values of
$\chi^{2}$.

\begin{figure}
\includegraphics[height=3in]{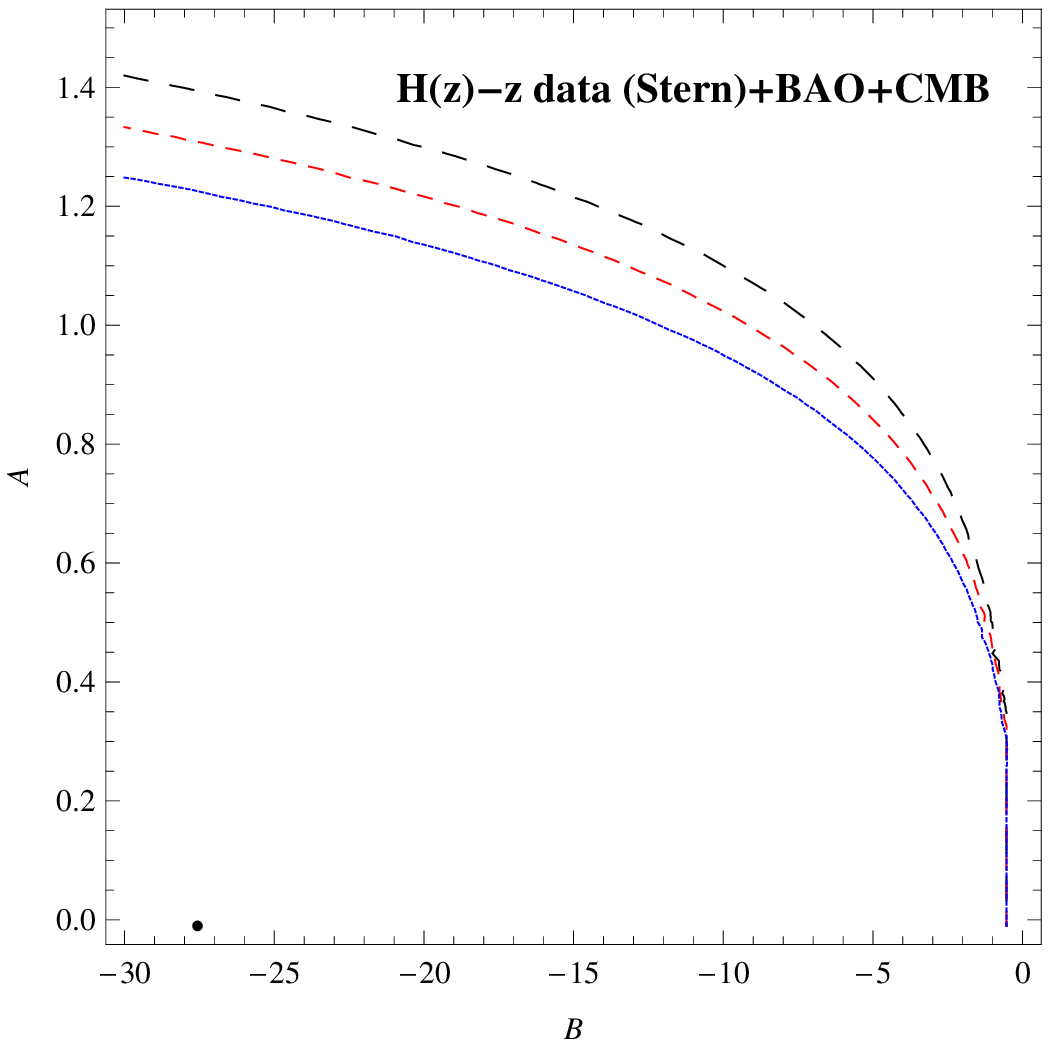}~~~~~\includegraphics[height=3in]{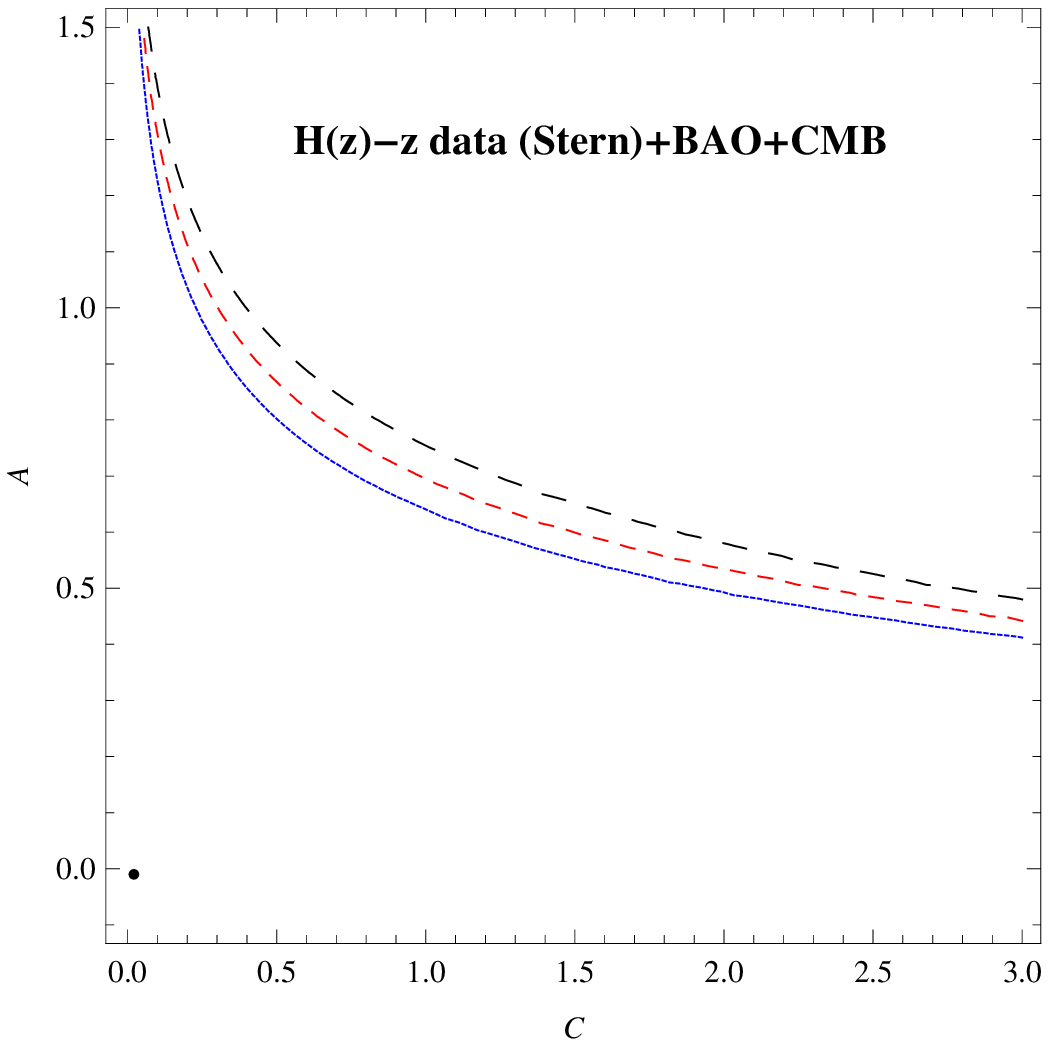}\\
\vspace{2mm}
~~~~~~~~~~~~~~~~~~~~~~~Fig.10~~~~~~~~~~~~~~~~~~~~~~~~~~~~~~~~~~~~~~~~~~~~~~~~~Fig.11\\
\vspace{1mm}

Fig.10 \& 11 show the variations of $A$ with $B$ and $A$ with $C$
respectively for different confidence levels in Galileon gravity
with  modified Chaplygin gas. The 66\% (solid, blue), 90\%
(dashed, red) and 99\% (dashed, black) contours are plotted in
these figures for the $H(z)$-$z$ (Stern)+BAO+CMB joint analysis.
\vspace{1cm}

\end{figure}

\subsubsection{Redshift-Magnitude Observations from Supernovae Type Ia}

The Supernova Type Ia experiments provided the main evidence for
the existence of dark energy. Since 1995, two teams of High-$z$
Supernova Search and the Supernova Cosmology Project have
discovered several type Ia supernovas at the high redshifts
\cite{Perlmutter, Riess, Riess1, Perlmutter1}. The observations
directly measure the distance modulus of a Supernovae and its
redshift $z$ \cite{Riess2,Kowalaski1}. Now, take recent
observational data, including SNe Ia which consists of 557 data
points and belongs to the Union2 sample \cite{Amanullah1}. From
the observations, the luminosity distance $d_{L}(z)$ determines
the dark energy density and is defined by

\begin{equation}
d_{L}(z)=(1+z)H_{0}\int_{0}^{z}\frac{dz'}{H(z')}
\end{equation}

and the distance modulus (distance between absolute and apparent
luminosity of a distance object) for Supernovas is given by

\begin{equation}
\mu(z)=5\log_{10} \left[\frac{d_{L}(z)/H_{0}}{1~MPc}\right]+25
\end{equation}

The best fit of distance modulus as a function $\mu(z)$ of
redshift $z$ for our theoretical model and the Supernova Type Ia
Union2 sample are drawn in figure 12 for our best fit values of
$A$, $B$ and $C$ for (Stern) + BAO + CMB joint analysis as
$A=-0.010$, $B=-27.563$, $C=0.020$, $\Omega_{m0}=0.0014$,
$\Omega_{x0}=0.0014$, $\alpha=0.001$, $n=0.5$,$m=10$, $\omega=-3$,
$w_m=0.03$, $f_0=0.01$, $\phi_0=0.01$. From the curves, we see
that the theoretical MCG model in Galileon gravity is in agreement
with the union2 sample data.

\begin{figure}
~~~~~~~~~~~~~~~~~~~~~~~~~~~~~\includegraphics[height=2.5in]{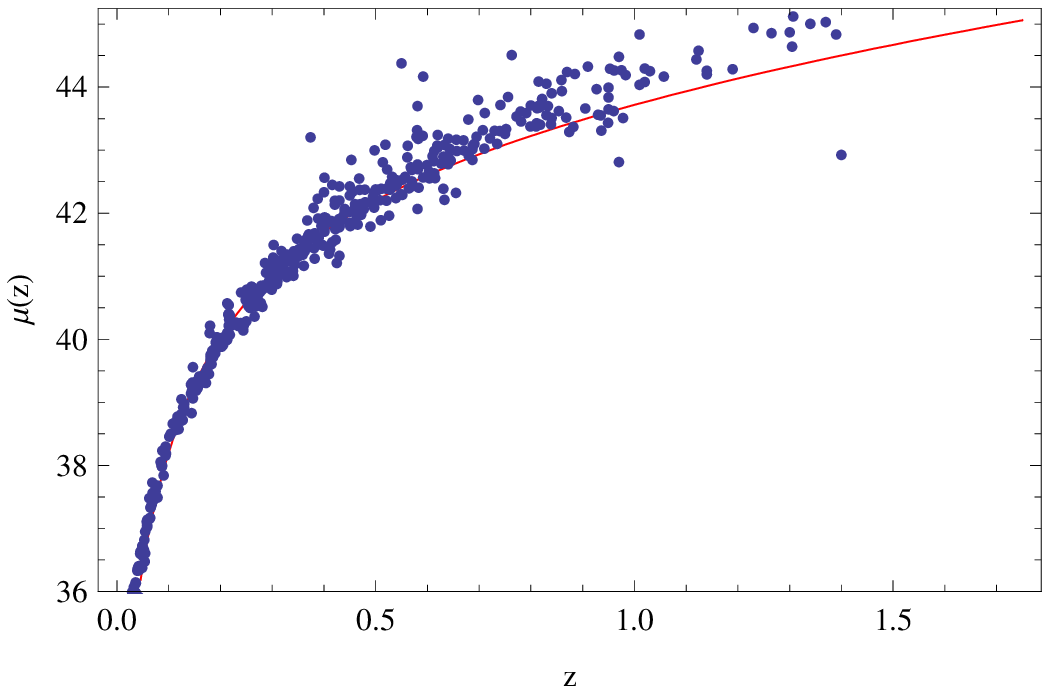}\\
\vspace{2mm}
~~~~~~~~~~~~~~~~~~~~~~~~~~~~~~~~~~~~~~~~~~~~~~~~~~~~Fig.12~~~~~~~~~~~~~~~~~~~~\\
\vspace{1mm}

Fig.12 shows $\mu(z)$ vs $z$ for Galileon gravity with  MCG (solid
red line) and the Union2 sample (dotted points). \vspace{1cm}

\end{figure}

\section{Discussions}

In this work, we have considered the FRW universe in Galileon
gravity filled with a combination of dark matter and dark energy
in the form of Modified Chaplygin gas (MCG). Since, MCG is one of
the candidate of unified dark matter-dark energy model. We present
the Hubble parameter in terms of the observable parameters
$\Omega_{m0}$, $\Omega_{x0}$ and $H_{0}$ with the redshift $z$ and
the other model parameters like $A$, $B$, $C$, $\alpha$ $n$, $m$,
$\omega$, $w_m$, $f_0$ and $\phi_0$. We have chosen the observed
values of $f_{0}=0.01, \phi_{0}=0.01, \alpha=0.001, w_m=0.03$,
$\omega=-3$, $n=0.5$, $m=10$, and $H_{0}$ = 72 Mpc$^{-1}$. In
figure 1, the plot of reconstructed deceleration parameter $q$ is
obtained against the redshift parameter $z$. It can be seen that
the present universe $(z=0)$ is undergoing a cosmic acceleration,
which is evident from the negative value of $q$. From the plot it
can also be predicted that this acceleration will continue late in
the future universe $(z<0)$. In figure 2, we have obtained the
plot of the EoS parameter ($w$) vs Redshift ($z$). In the curve
$z=0$ corresponds to $w=-0.5$ (approx.). Hence it is the testimony
of the fact that the current epoch is dominated by dark energy and
consequently accelerating in nature. It should also be noted that
the curve acquires an asymptotic behavior near $z=-1$, which
almost corresponds to $w=-1$. We know that physically the
admissible values of the redshift parameter is $z>-1$. For present
universe $z=0$ and for future universe $z<0$. In the interval
$-1<z<0$, the universe continues in the accelerating phase but,
the value of $w>-1$. This is true for almost any admissible values
of the model parameters. So it may be concluded that the
reconstructed model does not go beyond $\Lambda CDM$. In figures 3
and 4 the plots of the statefinder parameters ($r,s$) are
generated against redshift. These serves the purpose of
distinguishing the model under consideration with other models,
presenting a unique nature to it (the model). Finally in figure 5,
the trajectories in $r$-$s$ plane is obtained. It is worth
noticing that $r$ tends towards $1$ as $s$ tends towards $0$.
Therefore it is evident that these results tends towards the
$\Lambda CDM$ model.

From Stern data set (12 points), we have obtained the bounds of
the arbitrary parameters ($A,B$) \& ($A,C$) by minimizing the
$\chi^{2}$ test. Next due to joint analysis of \textit{Stern+BAO}
and \textit{Stern+BAO+CMB} observations, we have also obtained the
best fit values and the bounds of the parameters ($A,B$) \&
($A,C$). We have plotted the statistical confidence contour of
($A,B$) \& ($A,C$) for different confidence levels i.e.,
66\%(dotted, blue), 90\%(dashed, red) and 99\%(dashed, black)
confidence levels by fixing observable parameters $\Omega_{m0}$,
$\Omega_{x0}$ and $H_{0}$ and some other parameters $\alpha$, $n$,
$m$, $\omega$, $w_m$, $f_0$ and $\phi_0$ for \textit{Stern,
Stern+BAO} and \textit{Stern+BAO+CMB} data analysis.

From the \textit{Stern} data, the best-fit values and bounds of
the parameters ($A,B$) \& ($A,C$) are obtained and are shown in
Table 2 and the figures 6 \& 7 shows statistical confidence
contour for 66\%, 90\% and 99\% confidence levels. Next due to
joint analysis with \textit{Stern + BAO} data, we have also
obtained the best-fit values and bounds of the parameters ($A,B$)
\& ($A,C$) and are shown in Table 3 and in figures 8 \& 9 we have
plotted the statistical confidence contour for 66\%, 90\% and 99\%
confidence levels. After that, due to joint analysis with
\textit{Stern+BAO+CMB} data, the best-fit values and bounds of the
parameters ($A,B$) \& ($A,C$) are found and are shown in Table 4
and the figures 10 \& 11 shows statistical confidence contour for
66\%, 90\% and 99\% confidence levels. For each case, we compare
the model parameters through the values of the parameters and by
the statistical contours. From this comparative study, one can
understand the convergence of theoretical values of the parameters
to the values of the parameters obtained from the observational
data set and how it changes for different parametric values. The
distance modulus $\mu(z)$ has been drawn against redshift $z$ in
figure 12 for our theoretical model of the MCG in Galileon gravity
for the best fit values of the parameters of
\textit{Stern+BAO+CMB} data and the observed \textit{SNe Ia
Union2} data sample.

The observational study discovers the constraint of allowed
composition of matter-energy by constraining the range of the
values of the parameters for a physically viable MCG in Galileon
gravity model . We have also verified that when $\lambda$ is
large, the best fit values of the parameters and other results of
Galileon gravity model in MCG coincide with the results in
Einstein's gravity \cite{Paul}. When $\lambda$ is small, the best
fit values of the parameters and the bounds of parameters spaces
in different confidence levels in Galileon gravity model can be
clearly distinguished from Einstein's gravity for MCG dark energy
model. From the above discussion, we can conclude that the
observational data sets are perfectly consistent with our
predicted theoretical MCG model in Galileon gravity. Finally, it
is worth mentioning that, even though the quantum aspect of
gravity have small effect on the observational constraint, but the
cosmological observation can put upper bounds on the magnitude of
the correction coming from quantum gravity
that may be closer to the theoretical expectation than what one would expect.\\

\section*{Acknowledgements}

The authors sincerely acknowledge the facilities provided by the
Inter-University Centre for Astronomy and Astrophysics (IUCAA),
pune, India where a part of the work was carried out. Authors also
thank the anonymous referee for his/her invaluable comments that
helped them to improve the quality of the manuscript.\\


\begin{thebibliography}{99}

\bibitem{Perlmutter} Perlmutter, S. J. et al :- {\it Nature} {\bf 391} 51 (1998).
\bibitem{Riess} Riess, A. G. et al.[Supernova Search Team Collaboration] :- {\it Astron. J.} {\bf 116} 1009 (1998).
\bibitem{Riess1}  Riess, A. G. et al. :- {\it Astrophys. J.} {\bf 607} 665 (2004).
\bibitem{Bennet} Bennet, C. et al. :- {\it Phys. Rev. Lett.} {\bf 85} 2236 (2000).
\bibitem{Sperge} Spergel, D. N. et al. :- {\it Astrophys. J. Suppl. Ser.} {\bf 170} 377 (2007).
\bibitem{Adel}  Adelman-McCarthy, J. K. et al. :-  {\it Astrophys. J. Suppl. Ser.} {\bf 175} 297 (2008).
\bibitem{Eisenstein} Eisenstein, D. J. et al. [SDSS Collaboration] :-  {\it Astrophys. J.} {\bf 633} 560 (2005).
\bibitem{Briddle} Briddle, S. et al. :-  {\it Science} {\bf 299} 1532 (2003).
\bibitem{Spergel} Spergel, D. N. et al. :-  {\it Astrophys. J. Suppl.} {\bf 148}, 175 (2003).
\bibitem{Peebles} Peebles, P. J. E.,  Ratra, B. :- {\it Astrophys. J.} {\bf 325} L17 (1988).
\bibitem{Cald} Caldwell, R. R.,  Dave, R.,  Steinhardt, P. J. :- {\it Phys. Rev. Lett.} {\bf 80} 1582 (1998).
\bibitem{Arme} Armendariz - Picon,  C.,   Mukhanov, V. F.,  Steinhardt, P. J. :- {\it Phys. Rev. Lett.} {\bf 85} 4438 (2000).
\bibitem{Sen}  Sen, A. :- {\it JHEP} {\bf 0207} 065 (2002).
\bibitem{Cald1}  Caldwell, R. R. :-  {\it Phys. Lett. B} {\bf 545} 23 (2002).
\bibitem{Feng} Feng, B.,  Wang, X. L.,  Zhang, X. M. :- {\it Phys. Lett. B} {\bf 607} 35 (2005).
\bibitem{Kamen} Kamenshchik, A. Y.,  Moschella, U.,  Pasquier, V. :-  {\it Phys. Lett. B} {\bf 511} 265 (2001).
\bibitem{Debnath} Debnath, U., Banerjee, A. and Chakraborty, S., :-  {\it Class. Quantum Grav.} {\bf 21} 5609 (2001).
\bibitem{Cohen}  Cohen, A.,  Kaplan, D., Nelson,  A. :-  {\it Phys. Rev. Lett.} {\bf 82}, 4971 (1999).
\bibitem{Sahni}  Sahni, V.,  Shtanov, Y. :-  {\it JCAP} {\bf 0311} 014 (2003).
\bibitem{Cai} Cai, R. G. :-  {\it Phys. Lett. B} {\bf 657} 228 (2007).
\bibitem{Wei} Wei, H.,  Cai, R. G. :- {\it Phys. Lett. B} {\bf 660} 113 (2008).
\bibitem{Paddy1} Choudhury, T. R.,  Padmanabhan, T. :- {\it Astron. Astrophys.} {\bf 429} 807 (2007).
\bibitem{Tonry} Tonry, J. L. et al. :-  {\it Astrophys. J.} {\bf 594} 1 (2003).
\bibitem{Barris} Barris, B. J. et al. :- {\it Astrophys. J.} {\bf 602} 571 (2004).
\bibitem{Lu} Lu, J. et al. :- {\it Phys. Lett. B} {\bf 662} 87 (2008).
\bibitem{Jun} Dao-Jun, L.,  Xin-Zhou, L. :- {\it Chin. Phys. Lett.} {\bf 22} 1600 (2005).
\bibitem{Dvali} Dvali, G. R.,  Gabadadze, G.,  Porrati, M. :- {\it Phys. Lett. B} {\bf 484} 112 (2000).
\bibitem{An}  De Felice, A.,  Tsujikawa, T. :-  {\it arXiv}: 1002.4928 [gr-qc].
\bibitem{Noj0} S. Nojiri and S. D. Odintsov, arXiv:1011.0544 [gr-q].
\bibitem{clif}  Clifton, T.,  Barrow, J. :- {\it Phys. Rev. D} {\bf 72} 103005 (2005).
\bibitem{Yer} Yerzhanov, K. K. et al :- {\it arXiv}:1006.3879v1 [gr-qc] (2010).
\bibitem{Noj} Nojiri, S.,  Odintsov, S. D. :- {\it Phys. Lett. B} {\bf 631} 1 (2005).
\bibitem{An1} Antoniadis, I.,   Rizos, J.,  Tamvakis, K. :- {\it Nucl. Phys. B} {\bf 415} 497 (1994).
\bibitem{Hora}  Horava, P. :- {\it JHEP} {\bf 0903} 020 (2009).
\bibitem{Brans} Brans, C.,  Dicke, H. :- {\it Phys. Rev.} {\bf 124} 925 (1961).
\bibitem{Nicolis1} Nicolis, A., Rattazzi, R., Trincherini, E. :-  {\it Phys. Rev. D} {\bf 79} 064036(2009)
\bibitem{Ranjit1} Ranjit, C., Rudra, P., Kundu, S. :- {\it Astrophys. Space
Sci.} {\bf 347} 423(2013)
\bibitem{Chakraborty1} Chakraborty, S., Debnath, U., Ranjit, C. :-
{\it Eur. Phys. J. C.} {\bf 72} 2101(2012)
\bibitem{Debnath2} Debnath, U. :- {\it arXiv}:1310.2144 [gr-qc] (2013)
\bibitem{Deffayet1} Deffayet, C., Esposito-Farese, G., Vikman, A. :-  {\it Phys. Rev. D} {\bf 79} 084003 (2009)
\bibitem{Deffayet2} Deffayet, C., Deser, S., Esposito-Farese, G. :-  {\it Phys. Rev. D} {\bf 80} 064015 (2009)
\bibitem{Chow1} Chow, N., Khoury, J. :-  {\it Phys. Rev. D} {\bf 80} 024037 (2009)
\bibitem{Silva1} Silva, F. P., Koyama, K. :-  {\it Phys. Rev. D} {\bf 80} 121301 (2009)
\bibitem{Wu1} Wu, P. and Yu, H. :-  {\it Phys. Lett. B} {\bf 644} 16(2007)
\bibitem{Paul} Thakur, P., Ghose, S. and Paul, B. C.:-  {\it Mon. Not. R. Astron. Soc.} {\bf 397} 1935 (2009)
\bibitem{Paul1} Paul, B. C., Ghose, S. and Thakur, P., arXiv:1101.1360v1 [astro-ph.CO].
\bibitem{Sahni1} Sahni, V. et al.:-  {\it JETP} {\bf 77}, 201(2003).
\bibitem{Stern1} Stern, D. et al :-  {\it JCAP} {\bf 1002} 008 (2010).
\bibitem{Bond1} Bond, J. R. et al :- {\it Mon. Not. Roy. Astron. Soc.} {\bf 291} L33 (1997)
\bibitem{Efstathiou1} Efstathiou, G.,  Bond, J. R. :- {\it Mon. Not. R. Astro. Soc.} {\bf 304} 75 (1999)
\bibitem{Nessaeris1} Nessaeris, S.,  Perivolaropoulos, L. :- {\it JCAP} {\bf 0701} 018 (2007).
\bibitem{Komatsu1} Komatsu, E. et al. :- {\it Astrophys. J. Suppl.} {\bf 192} 18 (2011).
\bibitem{Perlmutter1} Perlmutter, S. J. et al :- {\it Astrophys. J.} {\bf 517} 565 (1999).
\bibitem{Riess2} Riess, A. G. et al. :-  {\it Astrophys. J.} {\bf 659} 98 (2007).
\bibitem{Kowalaski1} Kowalaski et al. :- {\it Astrophys. J.} {\bf 686} 749 (2008).
\bibitem{Amanullah1} Amanullah, R. et al. :- {\it Astrophys. J.} {\bf 716} 712 (2010).


\end{thebibliography}
\end{document}